\documentclass{article}

\usepackage{arxiv}

\usepackage[utf8]{inputenc} 
\usepackage[T1]{fontenc}    
\usepackage{hyperref}       
\usepackage{url}            
\usepackage{booktabs}       
\usepackage{amsfonts}       
\usepackage{nicefrac}       
\usepackage{microtype}      
\usepackage{lipsum}		
\usepackage{graphicx}
\usepackage[numbers]{natbib}
\usepackage{doi}
\usepackage{subcaption} 
\usepackage{color} 
\usepackage{adjustbox}
\usepackage{amsmath}
\usepackage{mathtools, nccmath}
\usepackage{xparse}
\usepackage{gensymb} 

\usepackage{multirow}
\usepackage{xcolor}
\usepackage{color,soul}
\usepackage{algorithm}
\usepackage{algorithmic}
\usepackage[normalem]{ulem} 
\usepackage{bm}
\usepackage{bbm}
\usepackage{amssymb}
\usepackage{accents}
\newcommand*{\dt}[1]{%
  \accentset{\mbox{\large\bfseries .}}{#1}}

\title{Optimization of pencil beam scanning pattern for FLASH proton therapy}
\usepackage{authblk}

\author[1]{Sylvain Deffet}
\author[1,2]{Edmond Sterpin}

\affil[1]{Molecular Imaging, Radiotherapy and Oncology, Institut de Recherche Expérimentale et Clinique (IREC), Université catholique de Louvain, 1200 Woluwe-Saint-Lambert, Belgium}
\affil[2]{Particle Therapy Interuniversity Center Leuven - PARTICLE, Leuven, 3000, Belgium}

\date{\today}

\renewcommand{\shorttitle}{FLASH dose rate}

\begin{document}
\maketitle

\begin{abstract}
\textbf{Background:} The FLASH effect, which reduces the radiosensitivity of healthy tissue while maintaining tumor control at high dose rates, has shown potential for improving radiation therapy. While the mechanisms behind the effect are not fully understood, it has been extensively studied with MeV electron beams and high-energy proton beams. However, to achieve FLASH proton therapy, changes to equipment and delivery systems are needed. Conformal FLASH proton therapy involves advanced beam-shaping technologies and specialized nozzle designs to confine the dose to the target volume. Optimizing the spot delivery pattern and range modulators can enhance the local dose rate, and genetic algorithms have been used to optimize scan patterns for stereotactic FLASH proton therapy of early-stage lung cancer and lung metastases.\\
\textbf{Purpose:} Maximize the dose rate within regions of interest through an efficient approach grounded in graph theory.\\
\textbf{Methods:} We have created a graph-based algorithm to optimize the trajectory of proton spots to maximize the 100th percentile dose rate. Since this problem is NP-hard, we have employed an approximation algorithm that can solve this kind of Traveling Salesman Problem (TSP) efficiently.\\
\textbf{Results:} When compared to a conventional serpentine pattern, the optimized scanning trajectory led to a doubling of the median dose rate, but only a minor increase in DR95. Our approach is more efficient and requires fewer evaluations of the objective function and hyper-parameters compared to existing genetic algorithms.\\
\textbf{Conclusions:} The optimized scanning trajectory led to a doubling of the median dose rate, but only a minor increase in DR95. The extent to which the dose rate can be increased depends on the size and shape of the region of interest. Future research could explore integrating FLASH objectives into treatment planning and incorporating the proposed method into plan optimization.
\end{abstract}


\section{Introduction}
The FLASH effect, which refers to a significant reduction of the radiosensitivity of healthy tissue while maintaining tumor control at ultrahigh dose rates, has garnered considerable interest in the radiation therapy community since it was reported in 2014\cite{Favaudon2014}. Although FLASH has a huge potential to increasing the therapeutic bandwidth of radiation therapy, the mechanisms behind the effect are still not fully understood, but possible explanations focus on the role of oxygen\cite{Favaudon2022}, radiochemistry, and the immune system\cite{Pratx2019,Spitz2019,LABARBE2020,JIN2020,Friedl2022,Favaudon2022}. The effect has been most extensively studied in radiobiological experiments with MeV electron beams, and a first patient has been successfully treated with FLASH electron beam therapy. Recent research has demonstrated that the FLASH effect is also present in high-energy proton beams\cite{Cunningham2021}, which may be particularly suited for deep-seated targets.

One of the main challenges in developing FLASH proton therapy is the need to significantly increase the dose rate. Preclinical experiments are typically conducted using small fields covered by a passive scattering method, but the maximum achievable field size for a given dose rate is directly limited by the maximum current that the system can output. However, the pencil beam scanning (PBS) technique has the potential to locally achieve high dose rates due to the limited size of the spots that are delivered individually.

To achieve a FLASH dose rate in PBS, a number of changes to the equipment and delivery system used in intensity modulated proton therapy (IMPT) are needed. In particular, the proton beam must be delivered at a much higher energy than is generally used in IMPT to ensure a high transmission efficiency\cite{JOLLY202071}. Additionally, using a treatment plan that involves multiple energies incurs delays required by the system to switch from one energy to the next, negatively impacting the average dose rate.

One approach to delivering a FLASH treatment involves shooting at maximum energy through the patient with so-called transmission beams, but this results in a significant amount of dose being delivered after the tumor, compromising the superior dosimetric potential of protons\cite{Rothwell2022}.

Conformal FLASH proton therapy, on the other hand, involves the use of advanced beam-shaping technologies and specialized nozzle designs to confine the dose to the target volume, similar to IMPT. A patient-specific range modulator, located between the nozzle and the patient, is used to tailor the range of the proton beam, enabling a conformal treatment plan with a single high-energy layer. Several methods have been proposed to optimize range modulators\cite{Simeonov2022, Liu2022, Zhang2022, Deffet2023}.

In addition to utilizing high energy beams, optimizing the spot delivery pattern can enhance the local dose rate\cite{JOSESANTO2022}. This optimization is a combinatorial problem that requires the use of approximation algorithms when the dimensionality of the problem increases. For instance, José Santo~\textit{et~al.}\cite{JOSESANTO2022} applied genetic algorithms to optimize scan patterns for stereotactic FLASH proton therapy of early-stage lung cancer and lung metastases. These algorithms can easily adapt to complex cost functions such as maximizing the dose rate in specific organs at risk. However, there are costs associated with these methods, such as the need to tune hyper-parameters and the execution time required.

We propose a fast and effective method to optimize the dose rate in specific regions of interest (ROIs) based on graph theory. In an \textit{in~silico} study, the optimized pattern is then compared with the conventional serpentine pattern for a head and neck case in terms of computed dose rates.

\section{Materials and Methods}
\subsection{Dose rate definition}
In order to optimize the dose rate, an essential first step is to establish a clear definition. In PBS proton therapy, local variations in dose rate occur as each voxel receives dose contributions from nearby PBS spots. Therefore, it is essential to consider the irradiation time of the spots and the time taken for the pencil beam to move from one position to the next. These factors have been incorporated into the definition of the PBS dose rate which has been established by Folkerts~\textit{et~al.}\cite{Folkerts2020, Varian2023}. This definition was later extended to explicitely include a dose threshold expressed as a percentage of the dose delivered to the voxel, resulting in the following percentile dose rate:\cite{Deffet2023b}:
\begin{equation}
    \dt{D}^{P}_i = \frac{pD_i}{t_{1, i} - t_{0, i}}
    \label{eq:perc_dr}
\end{equation}
where $t_{1,i}=t_i(\frac{(1-p)}{2}2D_i)$, $t_{0,i}=t_i(\frac{(1+p)}{2}D_i)$, $p$ is an arbitrary percentage, and $D_i$ it the total accumulated dose received in voxel $i$.

The maximum percentile dose rate proposed by Deffet~\textit{et~al.}\cite{Deffet2023b} extends the concept of percentile dose rate by considering all time windows in which the accumulated dose is at least $pD_i$:
\begin{eqnarray}
\dt{D}^{p}_i &=&\max_{t_0, t_1} \frac{\int_{t_0}^{t_1}d_i(t)dt}{t_1 - t_0}\label{eq:ODR}\\
    &s.t.& \int_{t_0}^{t_1}d_i(t)dt \geq p D_i \nonumber\\
    && t_1 > t_0 \nonumber
\end{eqnarray}
where  $d_i(t)$ is the dose received in voxel $i$ at time $t$.

In the present study, we focus on optimizing the 100-percentile dose rate for each voxel, which is the dose delivered to the voxel divided by the corresponding time interval over which it is delivered. Our selection of the 100-percentile dose rate as the optimization objective is the result of the graph representation that will be introduced later in the paper, and which will be thoroughly discussed.

According to Eq.~\ref{eq:perc_dr}, the 100-percentile dose rate is:
\begin{equation}
\dt{D}^{100}_i(\mathbf{S}) = \frac{D_i}{t_1(\mathbf{S}) - t_0(\mathbf{S})} = \frac{D_i}{T_i^{100}(\mathbf{S})} \label{eq:DR}\\
\end{equation}
where $\mathbf{S}$ is the spot sequence, $t_1(\mathbf{S}) = t_i(D_i^-)$ and $t_0(\mathbf{S}) = t_i(0^+)$ and where we introduce $T_i^{100}(\mathbf{S}) = t_1(\mathbf{S}) - t_0(\mathbf{S})$ as the quantity will be used many times later.

$\dt{D}^{100}_i(\mathbf{S})$ is equivalent to the PBS and percentile dose rates with a dose threshold of $0^+~\mathrm{Gy}$. The use of a threshold of $0^+~\mathrm{Gy}$ instead of $0~\mathrm{Gy}$ is the mathematical expression that we do not want to count the time when no dose is given before the first dose contribution to the voxel and also after the last contribution to the voxel.

\subsection{Optimization problem}
The objective is to maximize the dose rate in some specific ROIs, ie. to solve:
\begin{equation}
    \arg \max_{\mathbf{S}} \sum_{i \in ROI} \dt{D}^{100}_i(\mathbf{S})
\end{equation}

We relax this objective and rather consider the minimization of the time required to deliver the dose to each voxel:
\begin{equation}
    \arg \min_{\mathbf{S}} \sum_{i \in ROI} T_i^{100}(\mathbf{S})
\end{equation}
In other words, we are going to determine the spot sequence that minimizes the averaged time required to deliver the dose to each voxel. One very important point is that for each voxel, we do not count the time when no dose is given before any dose contribution to the voxel and also after that the voxel has received its full dose.

\subsubsection{Optimization of the delivery time of the whole field}
Our initial focus is on the fundamentals of graph theory as applied to the optimization of treatment field delivery time. While this is not our primary objective, this exercise is helpful in laying the groundwork for many of the definitions and concepts that will be utilized later on.

All the possibles delivery timing can be represented by means of a fully connected graph, named $G$. The corresponding adjacency matrix is called $\mathbf{M}$. $M_{s_1, s_2}$ is the time, also named $T_{s_1 \rightarrow s_2}$, required to move from spot $s_1$ to spot $s_2$. $M_{s_1, s_1}$ is the time, also named $T_{s_1}$, is the irradiation time associated to spot $s_1$:

\begin{equation}
\mathbf{M} =
\begin{bmatrix}
T_{s_1} & T_{s_1 \rightarrow s_2} & T_{s_1 \rightarrow s_3} & \dots\\
T_{s_2 \rightarrow s_1} & T_{s_2} & T_{s_2 \rightarrow s_3} & \dots\\
\vdots & \vdots & \ddots & \vdots \\
\dots & \dots & \dots & \dots
\end{bmatrix}\nonumber
\end{equation}

According to our simple delivery model, $M$ is symmetrical, i.e. $T_{s_1 \rightarrow s_2} =  T_{s_2 \rightarrow s_1}$ for all $s_1, s_2$.

In the context of optimizing the time required to deliver a treatment field, the eligible delivery sequences refer to all the possible routes that visit each spot exactly once. If a delivery sequence is feasible, the delivery time of the whole treatment field can be represented by the integrated distance along the sequence. Thus, we can optimize the time required to deliver the whole field by finding the shortest path in $G(\mathbf{M})$. This can be viewed as an instance of the well-known Travelling Salesman Problem (TSP), for which numerous algorithms exist in the literature, and which can be selected depending on the scale of the problem at hand.

An equivalent way to build such a graph is not to place the delivey times $T_s$ on the vertices but to add them on the edges. The corresponding adjancency matrix is thus:
\begin{equation}
\mathbf{M} =
\renewcommand\arraystretch{1.3}
\left[
\begin{array}{c|ccc}
0 & T_{s_1} & T_{s_2} & \dots\\
\hline
0 & 0 & T_{s_1 \rightarrow s_2} + T_{s_2} & \dots\\
0 & T_{s_2 \rightarrow s_1} + T_{s_1} & 0 \dots\\
\vdots & \vdots & \ddots & \vdots \\
0 &\dots & \dots & \dots
\end{array}
\right]
\label{matrix2}
\end{equation}
This matrix has zeros everywhere on the diagonal and is not symmetrical anymore. This matrix has one more row and colums with respect to the previous one. The first row ensures that the beam-on time of the first spot to be delivered will be accounted for. The  first column set to zero everywhere ensures that the first row can only be reached once.

In this representation, a notable advantage is that considering the shortest path on the corresponding graph results in the consideration of all possible starting spots simultaneously, due to the addition of an extra, fictional spot $s_0$. However, if a specific starting point is desired, the first row and column can be removed, and the matrix can be reorganized such that the starting spot occupies the first row and column.

\subsubsection{Optimization of the local dose rate}
In light of the above method for optimizing the delivery timing of the entire treatment field, we can now focus on the specific task of optimizing the sum of $T_i^{100}$ across all voxels within a designated region of interest (ROI).

Considering a sequence of spots $\mathbf{S}$ which is a path $P$ in the graph, the sum of $T_i^{100}$ can be obtained by computing, for each voxel, the minimum spanning subgraph in $G$ which spans over the spots which contribute to $i$:
\begin{eqnarray}
    \sum_i T_{i}^{100}(P, \mathbf{M}) &=& \sum_i P^{i}(\mathbf{M})\\
\end{eqnarray}
where $P^{i}$ is the minimal subpath in $P$ that delivers the full dose to voxel $i$, and $P^{i}(\mathbf{M})$ is the length of $P^{i}$ computed on $G(\mathbf{M})$.

To determine the optimal sequence we could use brute force:
\begin{enumerate}
    \item Compute every feasible paths $P$;
    \item For each feasible path:
    \begin{enumerate}
        \item For each voxel $i$, contribute the sets of spots which contribute to the dose of the voxel;
        \item Find the minimum spanning supbath $P^{sub} \in P$ which spans over all the spots which contribute to $i$ and compute its cumulative length $T_i$
        \item Add the length of this subpath to the cumulative sum $\sum_i T_{i, p}^{100}$
    \end{enumerate}
\end{enumerate}
The optimal path is the one that minimizes $\sum_i T_{i, p}^{100}$ over all possible paths $p$.

Given the infeasibility of using brute force to optimize problems with a significant amount of delivery spots, we propose a modified approach to graph optimization. To prioritize routes that contribute to the largest number of voxels which should decrease $\sum_i T_{i}^{100}$, we adjust the edge weights of the original graph used for optimizing the entire treatment field. Specifically, we construct an adjacency matrix $\mathbf{E}$, where $E_{s,s}=0$ for all $s$, and $E_{s_1,s_2}$ represents the ratio of (1) the time required to deliver a dose from spot $s_2$ when starting from spot $s_1$, and (2) the number of voxels in the ROI that receive a dose contribution from spot $s_2$.

For the sake of conciseness, we define the number of voxels that receive contributions from spots $s$ as follows.
\begin{equation}
    N_{s} = \sum_i \delta_{D_{i, s}>0}
\end{equation}

We now build the adjacency matrix $\mathbf{E}$ similarly to $\mathbf{M}$ but with
\begin{eqnarray}
    E_{s_1, s_2} &=& \frac{T_{s_1 \rightarrow s_2} + T_{s_2}}{N_{s_2}}\\
    &=& \frac{M_{s_1, s_2}}{N_{s_2}}\label{eq:denom}\\
\end{eqnarray}

The adjacency matrix $\mathbf{E}$ is thus:
\begin{equation}
\mathbf{E} = \mathbf{M}
\left[
\begin{array}{c|cccc}
1 & 0 & 0 & 0 & \dots\\
\hline
0 & \frac{1}{N_{s_1}} & 0 & 0 & \dots\\
0 & 0 & \frac{1}{N_{s_2}} & 0 & \dots\\
\vdots & \vdots & \vdots & \ddots & \vdots \\
0 &\dots & \dots & \dots & \dots
\end{array}
\right]
\label{matrixE}
\end{equation}

The proposed approach aims to favor transitions based on the amount of voxels that receive dose at the end of the transition.

The resulting algorithm for finding the optimal spot sequence is:
\begin{enumerate}
    \item Select the subsets $S$ of spots which contribute to the ROI
    \item Compute $T_{s_i} \forall s_i \in S$
    \item Compute $T_{s_i \rightarrow s_j} \forall s_i, s_j \in S$
    \item Compute dose influence matrix $D_{i, s} \forall i \in ROI, s \in S$
    \item From $D_{i, s}$, compute $N_{s_j} \forall s_j \in S$. The utilization of a threshold may be implemented to eliminate any contributions to the dose that are deemed not significant enough to warrant inclusion in the calculation.
    \item Compute $\mathbf{E}$ from $T_{s_i}, T_{s_i \rightarrow s_j}, N_{s_j}$
    \item Solve optimal\_ordering = TSP($\mathbf{E}$)
\end{enumerate}

\subsection{\textit{In silico} assessment}
We applied spot pattern optimization on conformal FLASH treatment plans calculated using the methodology presented in our prior publication\cite{Deffet2023}. In this paper, a treatment plan was optimized on a head and neck case which is reused in the present publication. The PTV considered in the present study had a prescription of 54~Gy. However, dose rates must be computed per fraction. As FLASH treatments will most likely be hypofractionated\cite{Van_de_Water2019-pk}, we considered that the dose per fraction would be around 8~Gy.

The treatment plan and its associated range modulator were computed for an energy of 226~MeV, a spot size of $(\sigma_x = 4.5~\mathrm{mm}, \sigma_y=5~\mathrm{mm})$. As the spot spacing has a direct impact on the scanning time and thus on the average dose rate, we considered two spot spacings: 4~mm which is close to that conventionnaly used, and 15~mm which is expected to yield higher dose rates. It is to be noted that because of the higher scattering introduced by the passive degradation of the beam, larger spot spacing may be used than in IMPT without introducing significant degradation of the dose\cite{Deffet2023b}.

To facilitate the comparison of the different dose rate formulas, we used a simple model where:
\begin{enumerate}
    \item the nozzle output current was considered constant;
    \item the time between 2 spots was proportional to the distance between the spots.
\end{enumerate}

In other words,
\begin{enumerate}
    \item the irradiation time of a spot was the ratio between the charge of the spot and the current;
    \item the time separating 2 spots was the ratio between the distance of the spots and an assumed scanning speed.
\end{enumerate}

The current was 500~nA (averaged on a pulse period) at the output of the nozzle and the scanning speed was 8000~mm/s. We considered that 1~MU corresponds to 152,880,000 protons.

A dose influence matrix was calculated using MCsquare\cite{Souris2016} to determine the contributions of each spot to the dose of each voxel.

In order to showcase the potential of the proposed approach, we applied it to optimize the spot pattern on both regions of interest (ROIs) corresponding to organs at risk as well as an extention of the PTV. The resulting dose rate maps were then compared to that obtained with the conventional serpentine pattern.

\section{Results}
\begin{figure*}
    \centering
    \includegraphics[width=\textwidth]{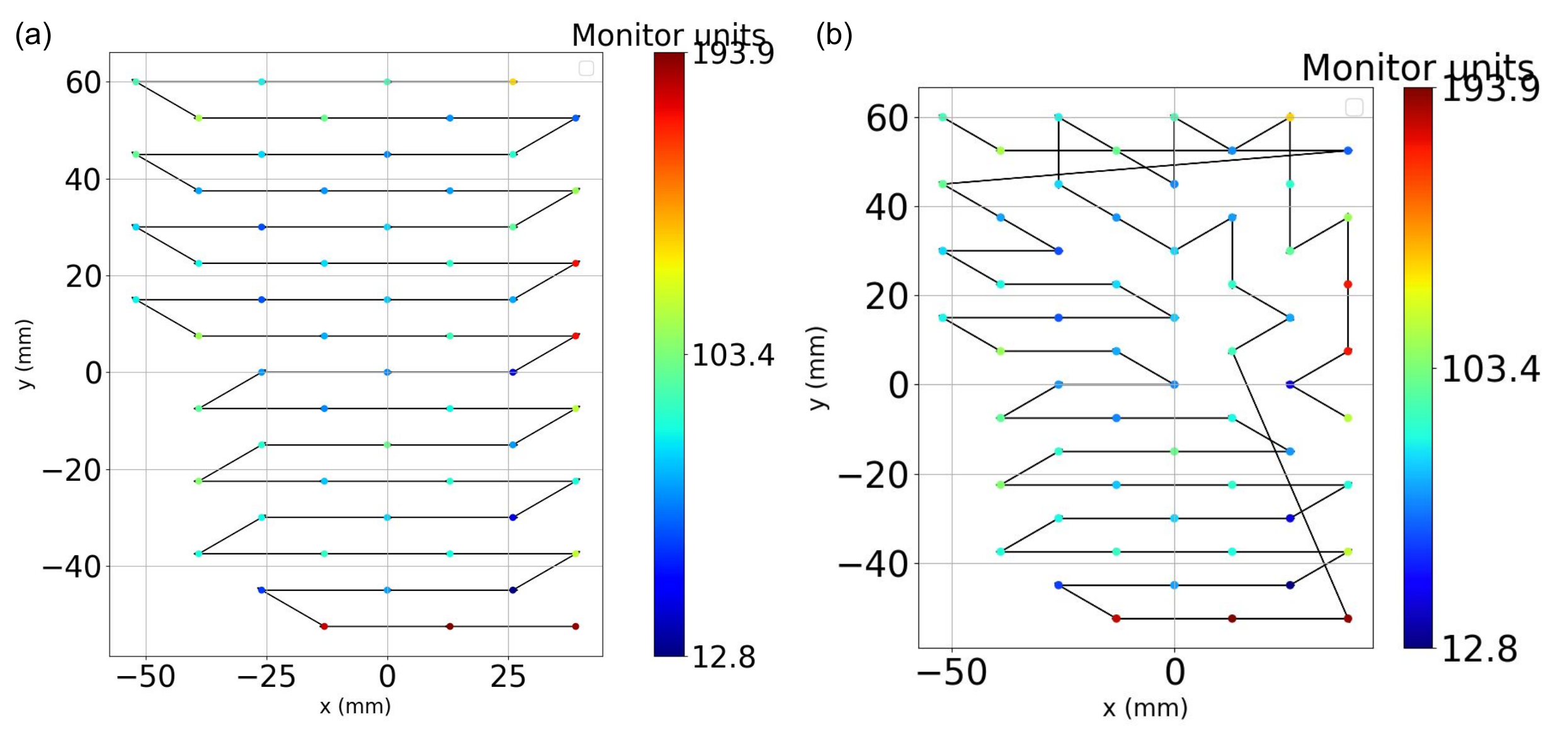}
    \caption{(a) Unoptimized spot pattern and (b) spot pattern optimized to maximize the dose rate in both the parotid and submadibular glands with a spot spacing of 15 mm.}
    \label{fig:spotPatterns}
\end{figure*}

\begin{figure*}
    \centering
    \includegraphics[width=\textwidth]{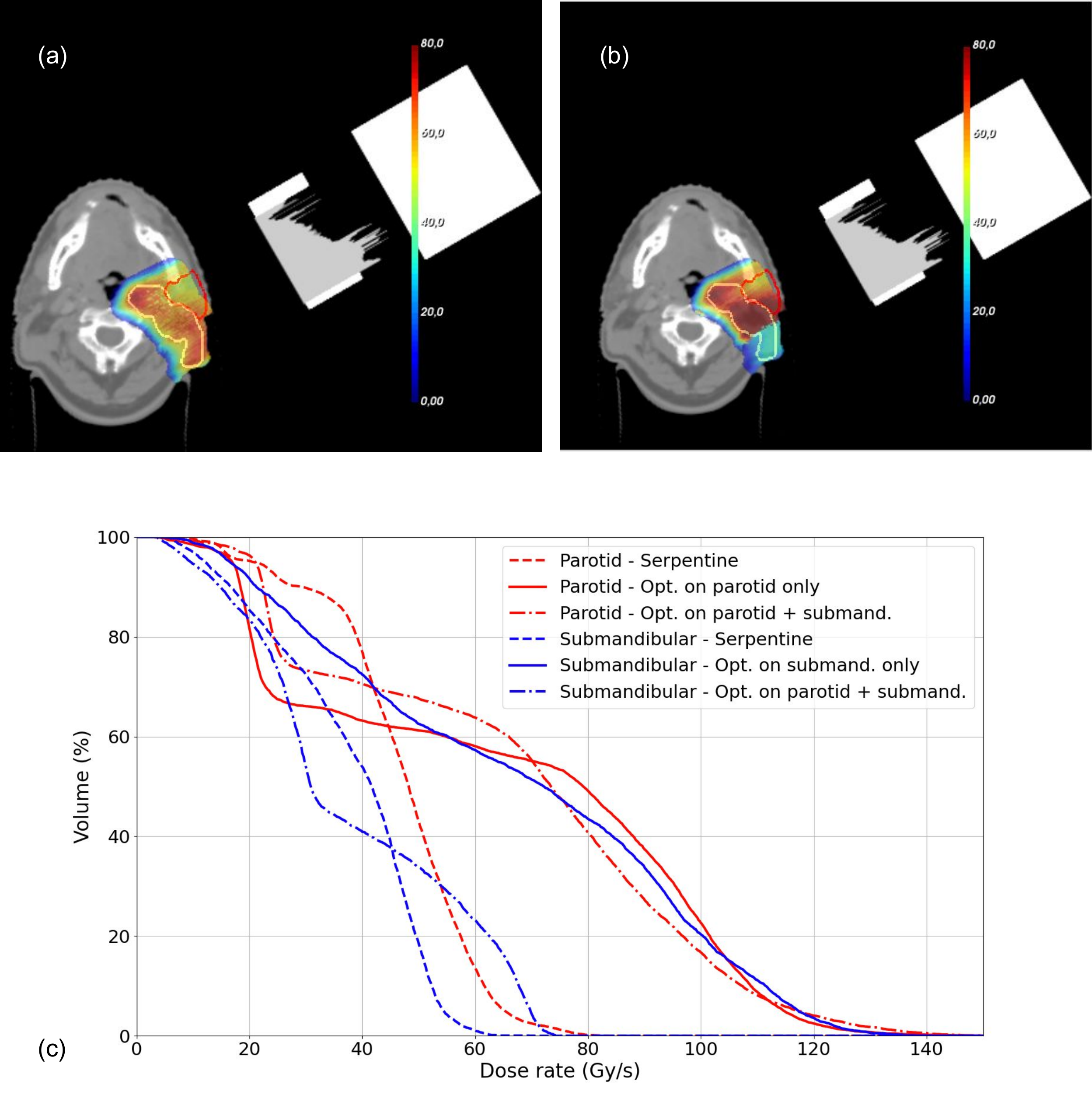}
    \caption{For a spot spacing of 15 mm, (a) 95-percentile dose rate for the unoptimized (serpentine) spot pattern and (b) for the spot pattern optimized to maximize the dose rate in both the parotid and submadibular glands. (c) 95-percentile dose rate volume histograms for unoptimized and optimized spot patterns.}
    \label{fig:parotid_submandibular}
\end{figure*}

In our study, the submandibular (brown contour) and parotid (red contour) glands in the head and neck region were used as examples of regions of interest in which we could optimize the dose rate. The submandibular gland, being proximal to the target volume, receives a significant dose, while the smaller parotid gland is mostly within the target volume. First, we optimized the dose rate exclusively in the submandibular gland using our algorithm. Then, we optimized the dose rate exclusively in the parotid. Finally, we sought to maximize the dose rate in both the parotid and submandibular glands using the algorithm. For each spot scanning pattern, we computed the 95-percentile dose rates and the corresponding DRVHs. The results are presented in Fig.~\ref{fig:parotid_submandibular} and Fig.~\ref{fig:spotPatterns} for a spot spacing of 15~mm. The dashed DRVHs represent the serpentine pattern and the solid DRVHs correspond to the optimized spot delivery pattern. The yellow DRVH in the figure corresponds to the PTV.

The results show a much higher DR50 with the optimized scanning pattern than with the conventional serpentine pattern. However, the DR95 is only slighty improved. In addtion, we see that the improvement is much higher when only one organ at risk is considered. Overall, this means that the improvement of the dose rate values is limited by the size and the complexity of the volume in which we want to maximize it, which is not suprising.

\begin{figure*}
    \centering
    \includegraphics[width=\textwidth]{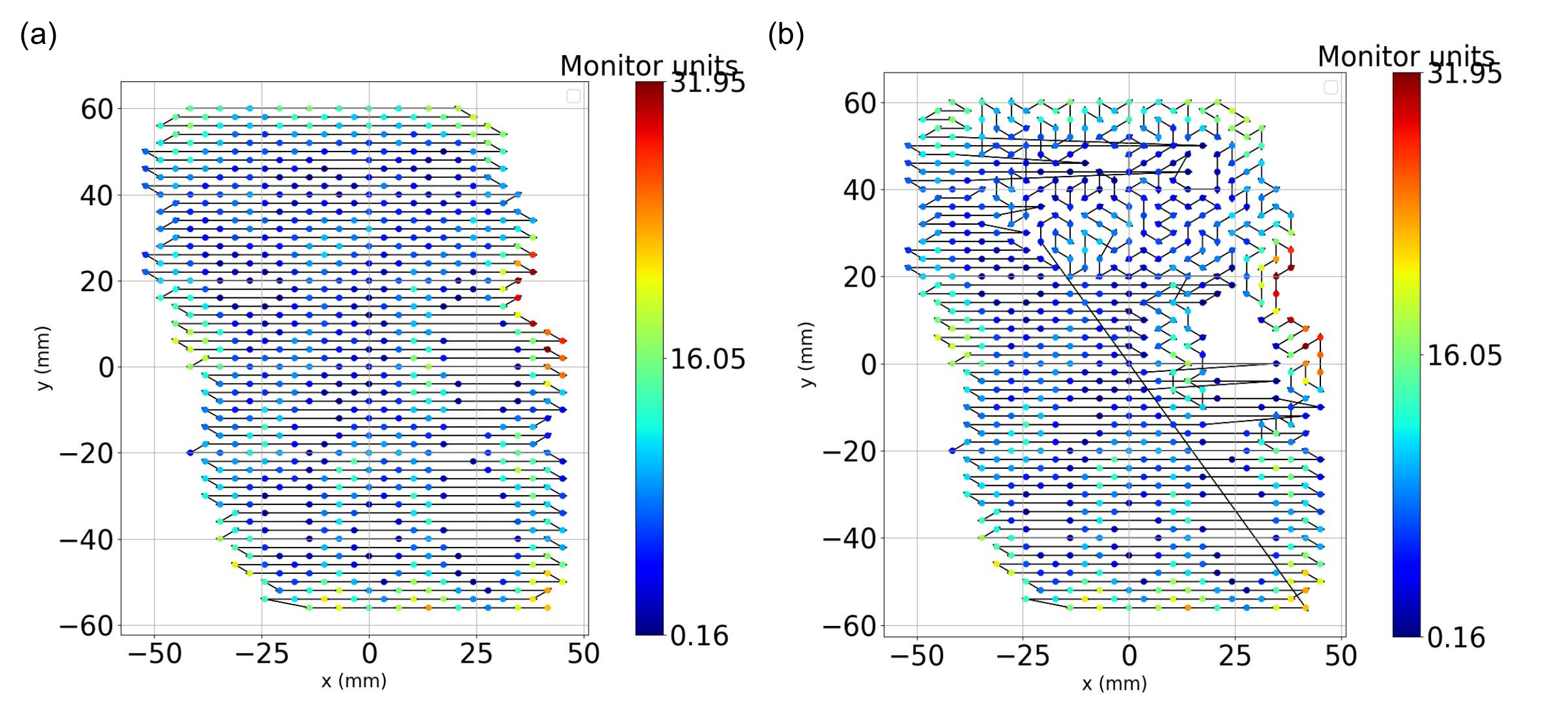}
    \caption{(a) Unoptimized spot pattern and (b) spot pattern optimized to maximize the dose rate in both the parotid and submadibular glands with a spot spacing of 4 mm.}
    \label{fig:spotPatterns_4mm}
\end{figure*}

\begin{figure*}
    \centering
    \includegraphics[width=\textwidth]{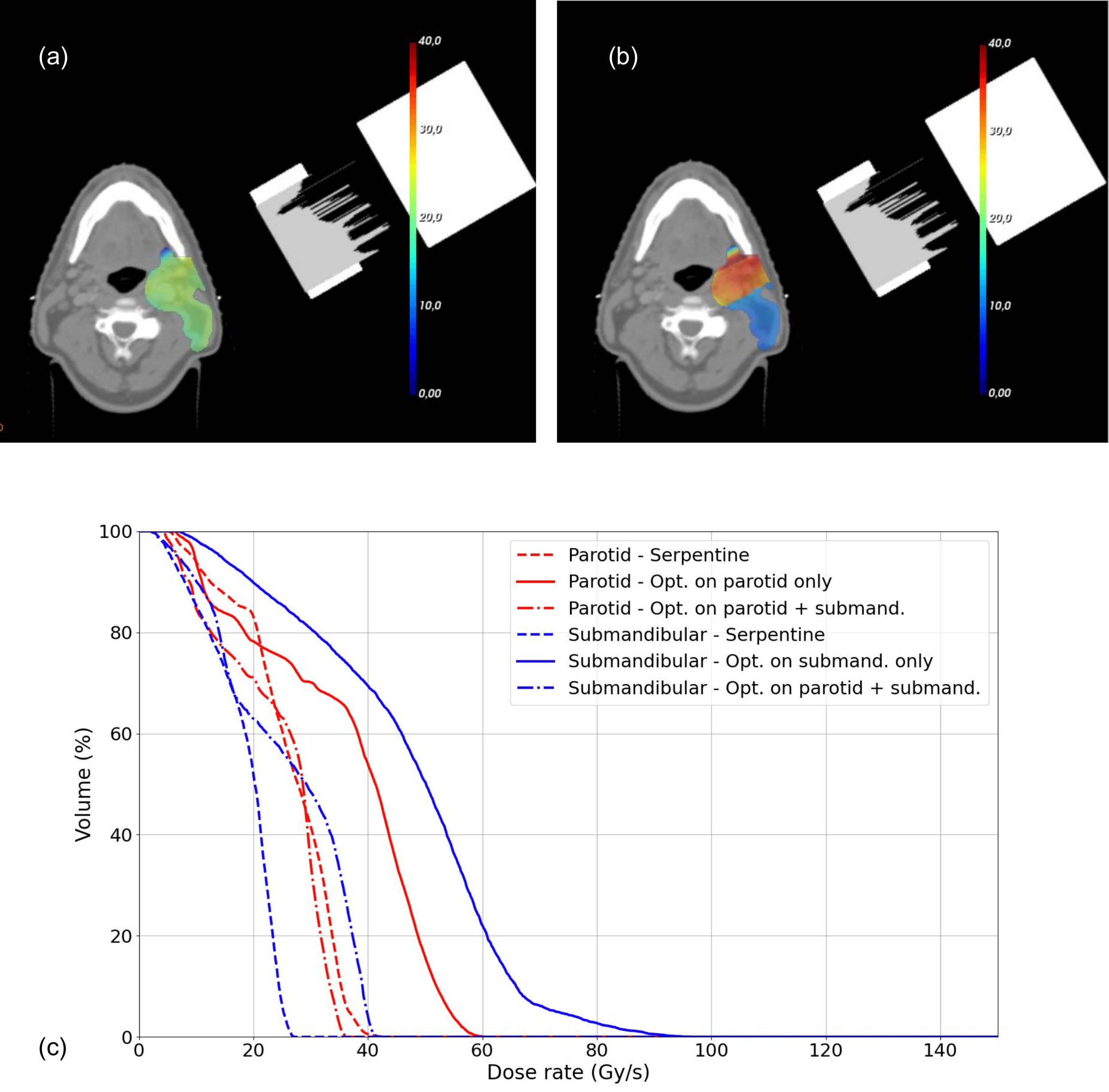}
    \caption{For a spot spacing of 4 mm, (a) 95-percentile dose rate for the unoptimized (serpentine) spot pattern and (b) for the spot pattern optimized to maximize the dose rate in both the parotid and submadibular glands. (c) 95-percentile dose rate volume histograms for unoptimized and optimized spot patterns.}
    \label{fig:parotid_submandibular_4mm}
\end{figure*}

We conducted the same optimization procedure on a treatment plan with a spot spacing of 4~mm. Fig.~\ref{fig:parotid_submandibular_4mm} shows that the resulting dose rate was lower than with a spot spacing of 15~mm which is due to the accumulated scanning time being larger as one can figure out by comparing spot maps in Fig.~\ref{fig:spotPatterns} and Fig.~\ref{fig:spotPatterns_4mm}. After optimizing the spot pattern, we found that a DR50 greater than 40~Gy/s was achievable in either the parotid or the submandibular gland, but not both simultaneously. These results demonstrate the crucial role that spot spacing plays in determining the maximum achievable dose rate.

\section{Discussion}
In PBS proton therapy, local variations in dose rate occur as each voxel receives dose contributions from nearby PBS spots. In order to accurately calculate the dose rate, it is essential to consider the irradiation time of a spot and the time taken for the pencil beam to move from one position to the next. These factors have been incorporated into the definition of the PBS dose rate, which we refer to as the percentile dose rate when the dose threshold is expressed as a percentage of the dose delivered to the voxel.

In our study, we have created a graph-based algorithm that optimizes the spot trajectory with the goal of maximizing the 100-percentile dose rate. It is widely acknowledged that solving such a graph-based problem is more efficient than utilizing genetic algorithms and involves fewer hyper-parameters to be fine-tuned. However, owing to the NP-hard characteristic of the TSP problem, approximation algorithms are typically necessary. In addition, it should be noted that genetic algorithms possess certain advantages too, such as modularity and the capability to optimize various definitions of the dose rate. For instance, the genetic algorithm proposed by José Santos~\textit{et~al.} could optimize both the 95- and 100-percentile dose rates. In contrast, it is not feasible to directly optimize the 95-percentile dose rate using our proposed method. Nonetheless, the 100-percentile dose rate can be considered the upper limit of the percentile dose rate, and optimizing it is expected to lead to an improvement in the 95-percentile dose rate too, as was observed in our \textit{in~silico} study.

In comparison to the serpentine pattern, our optimized scanning trajectory achieves a two-fold increase in the median dose rate. However, two important caveats should be noted. Firstly, the increase in DR95 is significantly lower than that of the median dose rate. Secondly, the degree to which the dose rate can be increased depends on the size and shape of the ROI being considered. It is unrealistic to expect that the simple optimization of the scanning trajectory can double the dose rate accross the entire CTV and its extension. Nevertheless, our results are promising with respect to organs at risk, which we prioritize to benefit from the FLASH effect. Integrating FLASH objectives into treatment planning and incorporating our proposed method into the optimization of the plan are potential avenues for future research.

\section{Conclusions}
A graph-based algorithm has been developed to optimize spot trajectory to maximize the dose rate in ROIs. The optimized scanning trajectory achieved a two-fold increase in the median dose rate but only a limited increase in DR95. The degree to which the dose rate can be increased depends on the size and shape of the region of interest, and integrating FLASH objectives into treatment planning and incorporating the proposed method into plan optimization are potential future research avenues.

\section{Acknowledgments}
This work was supported by the Walloon Region of Belgium through technology innovation partnership no. 8341 (EPT-1 – Emerging Proton Therapies Phase 1) co-led by MecaTech and BioWin clusters.

\section{Conflict of Interest Statement}
We have no conflicts of interest to disclose.

\bibliographystyle{plain}
\bibliography{biblio}

\end{document}